\documentclass[useAMS,usenatbib]{mn2e}

\usepackage[latin1]{inputenc}  
\usepackage{graphicx}
\usepackage{txfonts}

\title[Formic acid ionisation by solar wind particles]{Ionisation and dissociation of
cometary gaseous organic molecules by solar wind particles I: Formic
acid}
\author[S. Pilling et al.]{S. Pilling$^{1}$\thanks{E-mail:
spilling@lnls.br}, A. C. F. Santos$^{3}$, W. Wolff $^3$, M. M.
Sant'Anna$^{3}$, A. L. F. Barros$^{3}$ \and G. G. B. de Souza$^2$,
N. V. de Castro Faria$^3$ and H. M.
Boechat-Roberty$^{4}$  \\
\\
$^{1}$Laboratório Nacional de Luz Síncrotron, Caixa Postal 6192, CEP
13084-971, Campinas, SP, Brazil.\\
$^{2}$Instituto de Química, Universidade Federal do Rio de Janeiro -
UFRJ, Ilha do Fundão, CEP 21949-900, Rio de Janeiro, RJ, Brazil.\\
$^{3}$Instituto de Física, Universidade Federal do Rio de Janeiro -
UFRJ, Ilha do Fundão, Caixa Postal 68528, CEP 21941-972, Rio de
Janeiro, RJ, Brazil.\\
$^{4}$Observatório do Valongo, Universidade Federal do Rio de
Janeiro - UFRJ, Ladeira Pedro Antônio 43, CEP 20080-090, Rio de
Janeiro, RJ, Brazil.\\}
\begin{document}

\date{Received / Accepted}

\pagerange{\pageref{firstpage}--\pageref{lastpage}} \pubyear{2005}

\maketitle

\label{firstpage}


\begin{abstract} 
In order to simulate the effects of energetic charged particles
present in the solar wind colliding with the cometary gaseous formic
acid molecule (HCOOH), laboratory experiments have been performed.
The absolute ionisation and dissociation cross sections for this
molecule interacting with solar wind particles were measured
employing fast electrons in the energy range of 0.5 to 2 keV and
energetic protons with energies varying from 0.128 to 2 MeV. Despite
the fact that both projectiles lead to a very similar fragmentation
pattern, differences in the relative intensities of the fragments
were observed. Formic acid survives about 4-5 times more to the
proton beam than to the energetic electron collision.The minimum
momentum transfer in the electron impact case was estimated to be
3-38\% larger than the minimum momentum transfer observed with the
equivelocity protons. The UV photodissociation rates and half-lives
for HCOOH are roughly closer to the values obtained with energetic
electrons. It is consequently important to take electron impact data
into account when developing chemical models to simulate the
interplanetary conditions.

\end{abstract}

\begin{keywords} 
comets: general - molecular process - molecular data - Sun: solar
wind - astrochemistry
\end{keywords}

\section{Introduction}
To the present date, about 50 molecules have been detected in
comets. In particular, the glycine precursor molecule, formic acid
(HCOOH), has been detected in several comets including comet C/2001
Q4 (Neat), C/2004 Q2 (Machholz) (Biver et al. 2005) and comet C/1995
O1 (Hale-Bopp) (Bockelée-Morvan et al. 2000), which have shown a
formic acid abundance ([HCOOH]/[H$_2$O]) up to 0.09 \% at 1 AU from
the Sun . These cometary molecules are highly exposed to the solar
radiation (photons and particles) which induce several physical and
chemical reaction processes, like ionisation and molecular
dissociation.

Comets consist of blocks of ice, mainly H$_{2}$O, CO, and CO$_{2}$
molecules, and dust with a typical size of 0.1-10 km. According to
the Safronov theory most comets were formed in the Uranus-Neptune
region and were dynamically expelled to form the Oort cloud, a large
spherical cloud with a radius from $10^4$ to $10^5$ AU surrounding
the Sun (Safronov 1969; Mumma et al. 1993). The total number of
comets is approximatelly 10$^{12}$, and their expected total mass is
roughly 100 Earth masses. A much smaller number of comets, $\sim
10^9$, are located in the Kuiper belt at 50-500 AU (Weissman 1991).
Perturbations caused by the planets and nearby stars make some
comets move towards the inner solar system. Ices evaporate and pick
up dust at small heliocentric distances. Their flow is not
gravitationally bound, and effects of the solar radiation and wind
are responsible for the spectacular phenomena of the coma, dust and
plasma tails.

The atomic and molecular ions released by the dissociation
processess contribute to the alternative and efficient astrochemical
evolution process of complex molecular synthesis, since some
ion-molecule reactions do not have an activation barrier and are
also very exothermic (Largo et al. 2004, Woon 2002, Pilling et al.
2005). These simple ions interact further with other ions to produce
more and more complex molecules (Herbst \& Leung 1996).

The formic acid molecule has been also observed in several other
astronomical sources such as protostellar ices NGC 7538:IRS9,
condritic meteorites (Briscoe \& Moore 1993), dark molecular clouds
(Ehrenfreund \& Schuttle 2000; Ohishi et al. 1992 and references
therein) and regions associated with stellar formation (Zuckerman,
Ball \& Gottlieb 1971; Winnewisser \& Churchwell 1975; Liu et al.
2001, 2002). Fragmentation of the formic acid molecule by soft
X-rays, present in star-forming regions, has recently been studied
under laboratory conditions using synchrotron radiation
(Boechat-Roberty, Pilling \& Santos 2005).

This paper is the first in a series of experimental studies in an
attempt to simulate the interaction between electrons and protons
from the solar wind with cometary gaseous organic molecules. Formic
acid is not the most abundant molecule in the organic cometary
inventory. However, it plays an important role in the chemistry of
biomolecules like amino acids (Woon et al. 2002; Mendoza et al.
2004.). Crovisier et al. (2004) have calculated some upper limits
for the simplest amino acid, glycine (NH$_2$CH$_2$COOH), from radio
spectroscopic observations of comet C/1995 O1 (Hale-Bopp).

Two other reasons were also decisive for the choice of HCOOH for
this experimental study. First, if offers the possibility of making
a comparison between the interaction of energetic particles and
stellar X-rays photons with molecules, as previously studied
(Boechat-Roberty, Pilling \& Santos 2005). The second reason is the
low mass resolution of the time-of-flight system used in the proton
impact experiment, which prevents a reliable mass spectrum
measurement of highly hydrogenated molecules. We plan to circumvent
this limitation in the near future, making feasible studies
involving hydrogen-rich and abundant molecules like methane
(CH$_4$), methanol (CH$_3$OH), ethane (CH$_3$CH$_3$), ethanol
(CH$_3$OH), dimethyl-ether (CH$_3$OCH$_3$), acetic acid
(CH$_3$COOH), acetaldehyde (CH$_3$CHO) and acetonitrile (CH$_3$CN)
among others.

A short review of solar wind properties is presented in the next
section along with a discussion on the electron and proton flux. In
the following sections, we present the experimental results on the
ionisation and fragmentation of the formic acid molecule upon
interaction with fast electrons and protons. The two experimental
setups which have been used are described in section 3. In section
4, we present the results obtained for impact of both energetic
particles with the gaseous formic acid including the kinetic energy
released by ionic fragments, the momentum exchange with the
projectiles, and the determination of the absolute cross sections,
dissociation rates and half-lifes. In section 5, final remarks and
conclusions are given.

\section{Solar wind composition and interaction with cometary gas}

The solar wind (or even the stellar wind) is formed by the
hydrodynamical expansion of the  outer shell of the solar
atmosphere, the corona. This rarefied plasma region which is exposed
to the strong solar gravity, is permeated by magnetic fields. The
corona is heated by magnetic hydrodynamical wave (MHD) propagation,
reaching temperatures of about $2 \times 10^6$ K. Due to magnetic
reconnections and annihilations, energetic charged particles like
electrons, protons and alpha particles (He$^{++}$) are ejected from
the solar corona and interact with planetary atmospheres and comets.
The interaction between the solar wind and atomic and molecular
medium (interplanetary matter) has drastic consequences, beyond the
direct ionisation and dissociation process. Processes like charge
transfer, X-ray fluorescence and bremsstrahlung emission also occur
as described in detail by Krasnopolsky, Greenwood \& Stancil (2004)
and references therein.

The mean solar wind velocity of 450 km/s corresponds to ion energy
of $\sim$ 1000 eV/u (energy/atomic mass unit). The solar wind at
high heliographic latitudes is emitted from coronal holes, and its
velocity reaches roughly 750 km/s, corresponding to an energy of
$\sim$ 3000 eV/u. The ion velocity decreases in the comet from the
bow shock to the lower density regions, according to the Giotto
measurements in the comet Halley (Goldstein et al. 1987). The
elementar composition of the solar wind is generally similar to the
solar composition (Grevesse \& Sauval 1998). Relative element
abundances in the slow ($\sim$ 400 km/s) and fast ($\sim$ 750 km/s)
solar wind and in the solar photosphere are presented in
Krasnopolsky, Greenwood \& Stancil (2004) and von Steiger et al.
(2000). The mean He/H ratio in the solar wind varies between 0.04 -
0.08, originating from different sources.

In Table~\ref{tab:Solarwind} we present some solar wind averaged
values taken from Zirin (1988), Toptygin (1985) and Kroll \&
Trivelpiece (1973). Near Earth, the velocity of the solar wind
varies from 100-900 km/s. The average velocity is 450 km/s.
Approximately 800 kg/s of material is lost by the Sun as ejected
solar wind, a negligible amount compared to the Sun's light output,
which is equivalent to about $4.5 \times 10^9$ kg of mass converted
to energy every second. The wind is believed to extend out to
between 100 and 200 AU.

Measurements of the solar wind ion density (mainly H$^+$, He$^+$ and
He$^{++}$), as a function of distance from comet 1P/Halley, were
performed by the Giotto spacecraft (Fuselier et al. 1991; Shelley et
al. 1987), and by the Plasma Experiment for Planetary
\begin{table}
\centering \caption{Averaged Properties of solar wind at 1 AU
(adapted from Zirin 1988, Toptygin 1985 and Kroll \& Trivelpiece
1973).}
\setlength{\tabcolsep}{2pt} %
\label{tab:Solarwind}
\begin{tabular} {l l l}
\hline \hline
Properties             & Quiet times                       & Disturbed times \\
\hline
Density$^\dag$        &  $\sim$ 10 ions/cm$^{3}$           & 20-40 ions/cm$^{3}$ \\
Bulk speed            &  $\sim$ 450 km/s (100-600 km/s)    & $\sim$ 750 km/s (700-900 km/s) \\
Ion temperature       &  $\sim 8 \times 10^4$ K            & $\sim 3 \times 10^5$ K \\
Proton energy         &  $\sim 0.6$ keV                    & $\sim 3$ keV (1-10$^4$ keV) \\
Electron energy       &  $\sim 0.3$ eV  (0.1-10$^4$ eV)    & $\sim 1.5$ eV (0.1-10$^4$ eV) \\
Magnetic field        & 3 - 8 $\times 10^{-5}$ G     & 10 - 30 $\times 10^{-5}$ G   \\
Energy flux           & $\sim$ 0.5 erg/cm$^{2}$        & $\sim$ 15 erg/cm$^{2}$ \\
 \hline \hline
\multicolumn{3}{l}{$^\dag$ 95\% H$^+$, 4\% He$^{++}$ and traces of C, N, O, Ne, Mg, Si and Fe ions}\\
\end{tabular}
\end{table}
Exploration (PEPE) aboard the Deep Space 1 fly by of the comet
19P/Borrelly (Young et al. 2004). Those data have shown that the
solar wind ions also capture electrons from cometary gases and
induce some charge exchange processes with cometary gases like
H$_2$O and CO$_2$ (Greenwood et al. 2000).

\begin{figure}
 \centering
 \resizebox{\hsize}{!}{\includegraphics{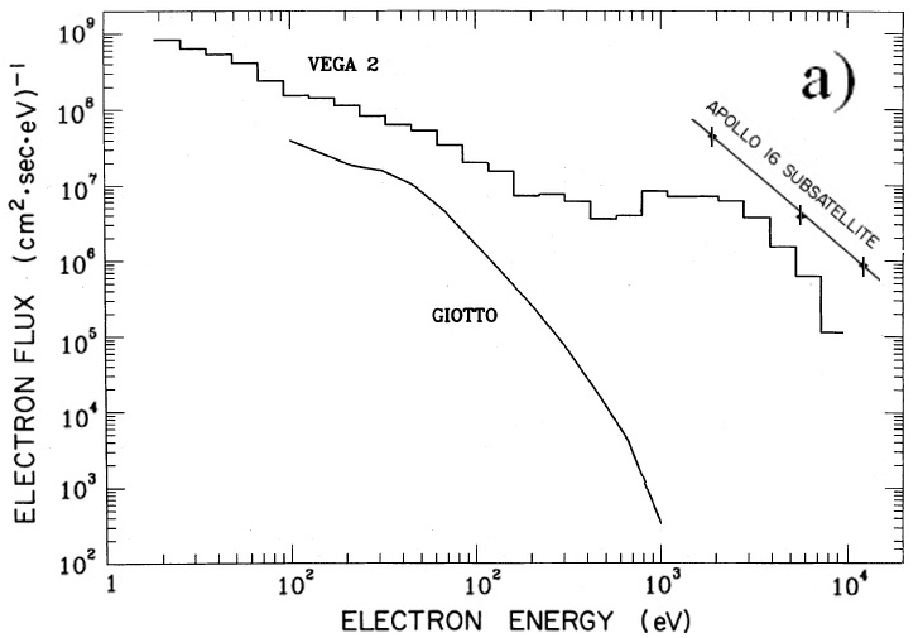}}
 \resizebox{\hsize}{!}{\includegraphics{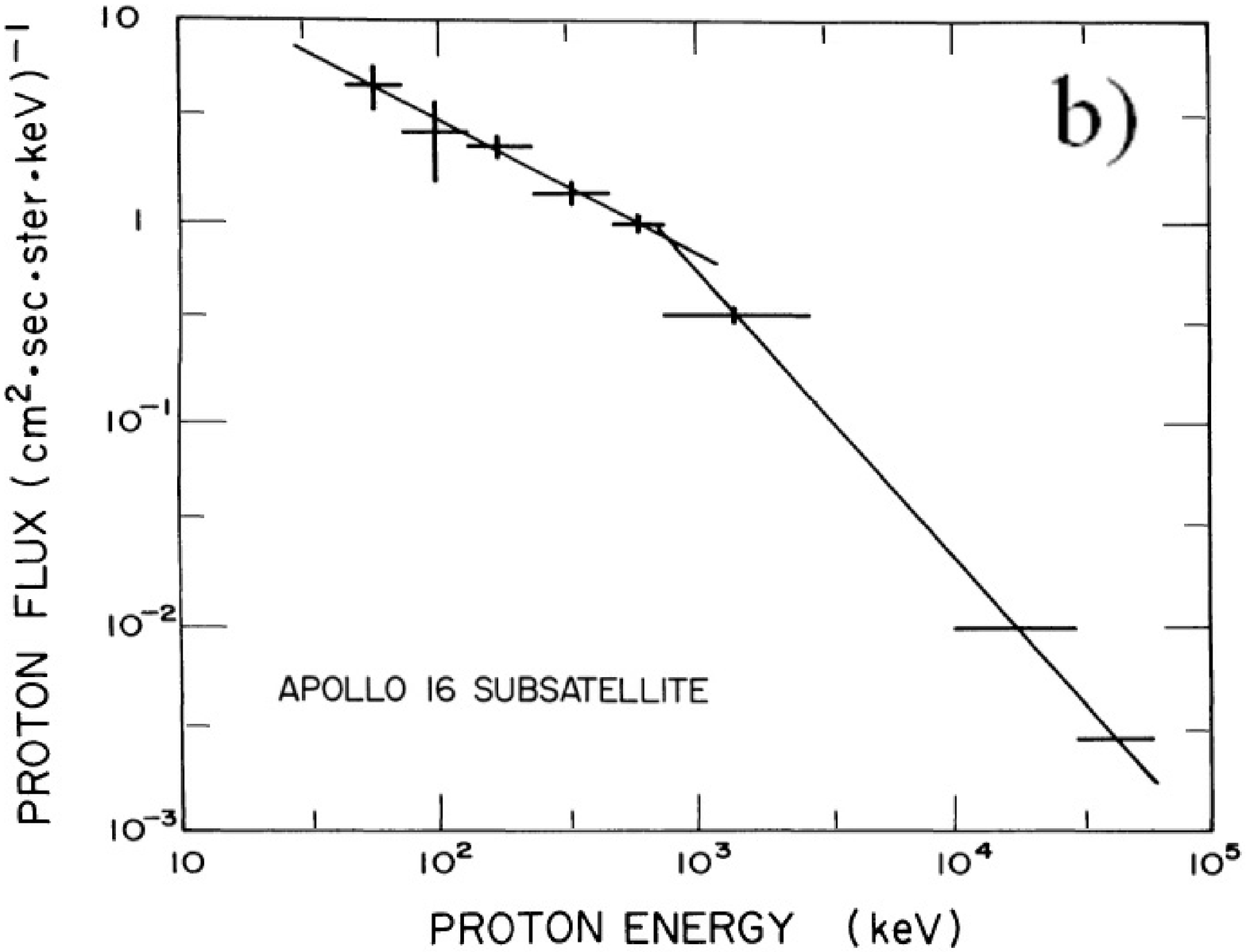}}
\caption{a) Electron fluxes in comet Halley measured by the Vega 2
(Gringauz et al., 1986) and by Giotto (d'Uston et al., 1989) and at
lunar orbit measured by APOLLO 16 subsatellite (adapted fron Lin,
McGuire and Anderson, 1974). b) The proton energy spectrum due to
solar wind at lunar orbit measured by APOLLO 16 subsatellite (Lin,
McGuire and Anderson, 1974)}
 \label{fig:particlesincomet}
\end{figure}

Depending on the cometary gas production rate and solar wind
conditions (and heliocentric distance), a bow shock or bow wave is
generated in the mass-loaded solar wind, preceding it, slowing
therefore the solar wind further, diverting the plasma flow around
the dense inner coma, and heating the streaming plasma. In addition,
a compression wave is formed. As the flow penetrates deeper into the
coma, the neutral gas density becomes sufficiently high so that some
ion-neutral particle collisions begin to dominate the fluid and the
chemical characteristics of the flow (Härberli et al. 1995). The
region where this occurs is termed collisionpause. Inside this
region (roughly $\sim 10^5$ km from the nucleus at Halley) two- and
even three-body collisions begin to take place, setting off a wide
range of complex physical and chemical process, including charge
transfer, ion-neutral collisions and electron impact ionisation
(Cravens 1997; Härberli et al. 1997). The charge transfer reduces
the ionised component of the solar wind which presents a maximum
where the ion mean free path begins to decrease rapidly (Gombosi
1987). Ion-neutral collisions lead to momentum exchange (Haberli et
al. 1995, 1997) and formation of the collisionpause roughly at the
point where the solar wind mean free path is comparable to the
distance of the comet nucleus (Cravens 1997).

According to Rodgers \& Charnley (2001) complex organic molecules
cannot be synthesized in the coma and should originate from the
nucleus. At lower altitudes, however, ionisation and dissociation
processes may occur and these energetic radicals may react with each
other or with the surrounding and abundant simple molecules (water,
CO, CO$_{2}$), promoting a slight increase in the molecular
complexity of the region. This scenario may also have a small
contribution on the formation of pre-biotic molecules, like the
amino acids, since one of their precursor, the carboxyl radical
(COOH) is produced in the fragmentation of formic acid, as we will
see further.

In Fig.~\ref{fig:particlesincomet}a we present the electron energy
distribution (taken from Krasnopolsky, Greenwood \& Stancil 2004) in
comet Halley, measured at various distances from the nucleus by the
Vega (Gringauz et al. 1986) and Giotto (d'Uston et al. 1989) probes
together whit the one obtained at lunar orbit measured by APOLLO 16
subsatellite (adapted from Lin, McGuire \& Anderson 1974). The
proton energy spectrum due to solar wind at lunar orbit measured by
APOLLO 16 subsatellite (Lin, McGuire \& Anderson 1974) can be seen
at Fig.\ref{fig:particlesincomet}b.

The mean energy of the solar wind protons is in the 1-2 keV range.
The K-shell tightly bound electrons of the C and O atoms, for
example, can be knocked off by solar wind protons. However, these
K-shell ionisation cross sections are extremely low and their
probability is much smaller than the fluorescence yield produced by
solar x-rays.

\section{Experimental setup}

The formic acid gas target is first degassed through freezing,
pumping and thawing cycles and then delivered in vapor form, through
a thin needle, into the interaction region  (volume defined by the
gas intersection between the effusive jet and the projectile beam is
about 1-3 mm$^3$). The base pressure in the vacuum chamber was in
the $10^{-8}$-$10^{-6}$ Torr range and during the experiment the
chamber pressure was maintained below $10^{-5}$ Torr. The pressure
at the interaction region was estimated to be $\sim$ 1 Torr
(10$^{16}$ mols cm$^{-3}$). These pressure values ensure a single
collision regime. The measurements were made at room temperature.

\subsection{Proton impact}

A tightly collimated monoenergetic proton beam with energies of
0.128, 0.2, 0.5, 1.0, 1.5 and 2.0 MeV is delivered by the 1.7 MV
Tandem accelerator of the Federal University of Rio de Janeiro. The
beam is charge analyzed by an electric field placed just before
crossing, at right angles, an effusive jet of formic acid molecules.
This is made in order to improve the charge state purity, since the
proton beam can become partially neutralised due to the interactions
with the residual gases present in the beam line. The emergent beam
is recorded by a channeltron detector housed in a detection chamber
downstream of the gas cell. A secondary electric field is applied
after the intersection with molecular beam in order to separate the
H$^0$ beam that performed single electron capture in interactions
with the gas target (see Fig. 2b). For the higher projectile
velocities, the direct ionisation process (the charge of the
projectile does not change during the collision) is the dominant
collision channel in the target ionisation, (Rudd et al. 1985a and
1985b). The average experimental proton flux was about 3 $\times$
10$^5$ protons cm$^{-2}$ s$^{-1}$.

\begin{figure}
 \centering
 \includegraphics[angle=0,scale=0.35]{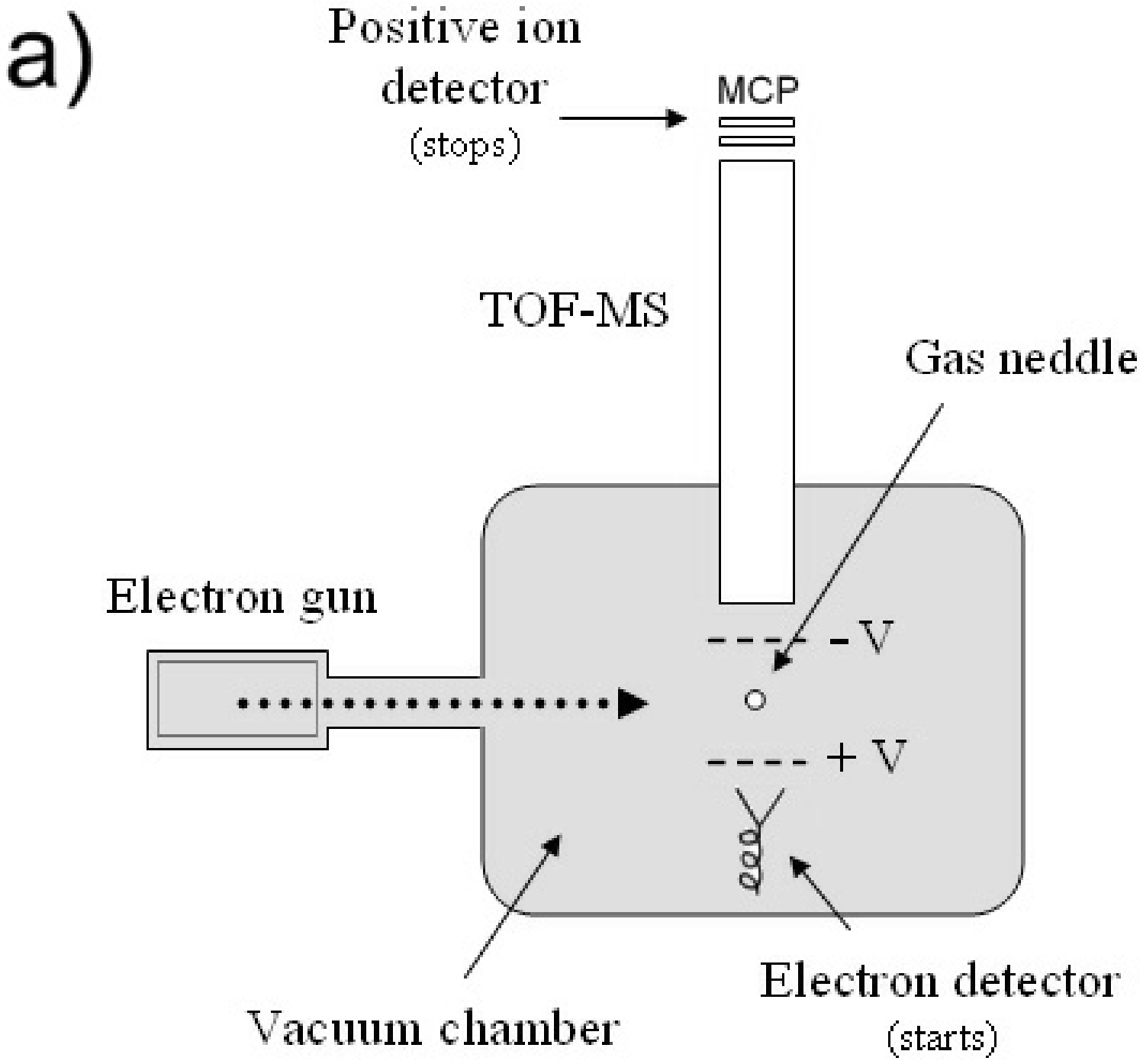}
 \includegraphics[angle=0,scale=0.35]{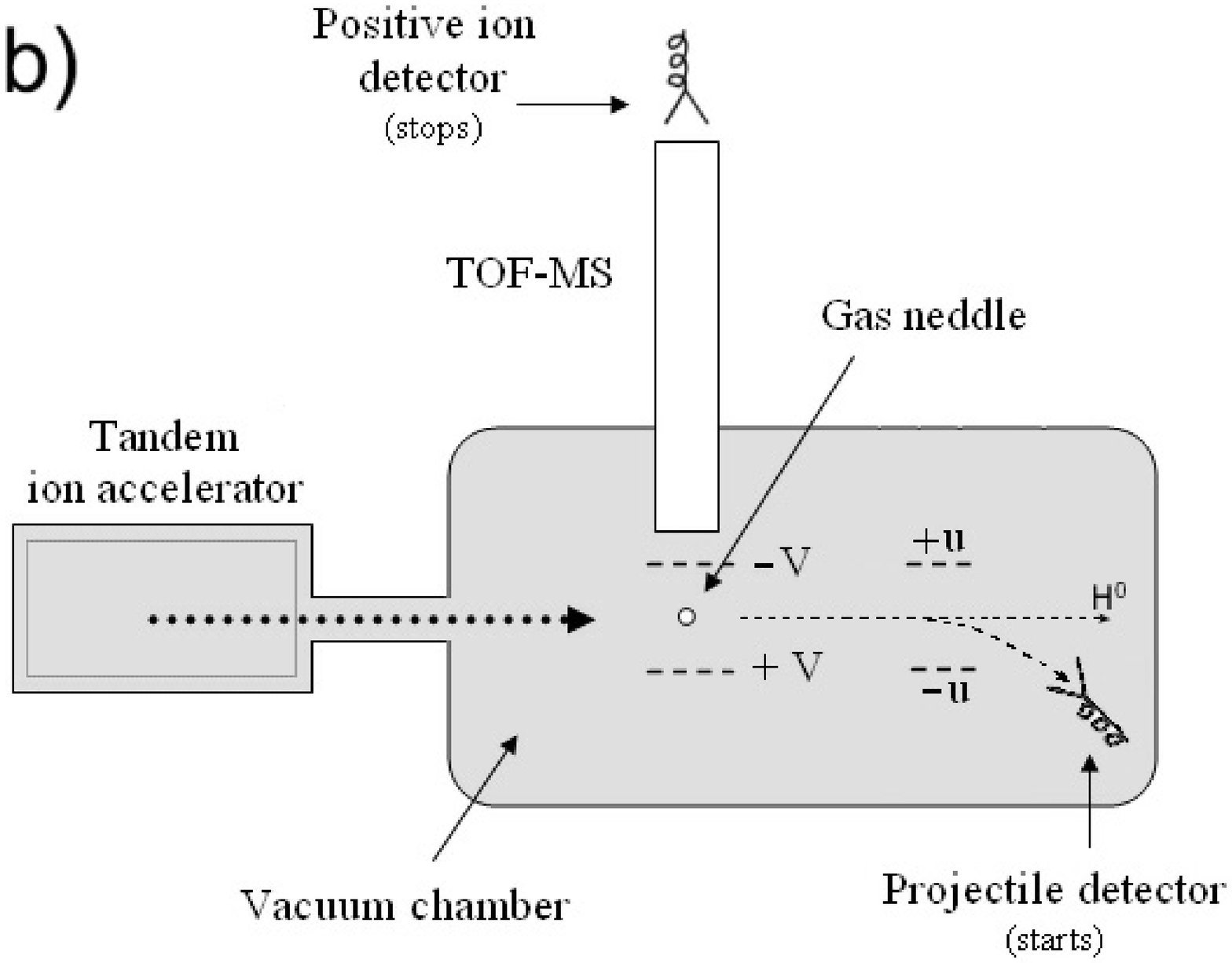}
 \caption{Experimental setup. a) high energy electron impact vacuum chamber. b) proton impact vacuum chamber.}
 \label{fig:TOFcartoon}
\end{figure}

The atomic and molecular ionic fragments ejected due to the
interaction with the incident beam are accelerated by a two-stage
electric field and detected by a channeltron detector. The time-of
flight (TOF) spectrometer, running in the coincidence mode, is
started by the detection of the proton beam and stopped by the ionic
fragments extracted onto the TOF-detector.

The first stage of the electric field is produced by a plate-grid
system subjected to a 130 V/cm electric field. The extracting
electric field is raised up to 950 V/cm in order to assure the
complete recoil collection. No differences are observed in the
recoils yield, indicating that most fragments are ejected with small
momenta. The limited mass resolution of the time-of-flight
spectrometer does not allow for the separation of ionic fragments
differing by 1 a.m.u.

\subsection{Electron impact}

The electron impact apparatus has been described in details by
Maciel, Morikawa \& de Souza (1997). Briefly, a D.C. monoenergetic
0.5, 1.0 and 2.0 keV electron beam is produced and crossed with the
gaseous formic acid target jet. The fragments resulting from the
collision are accelerated by a two-stage electric field and detected
by a microchannel plate detector mounted in a chevron configuration.
This detector provides the stop signals to the time-to-digital
converter started by the signals from the secondary low energy
electrons accelerated in the opposite direction. The first stage of
TOF-detector consists of a plate-grid system with the primary beam
passing through its middle with a 60 V/cm electric field. Ions with
energies up to 5 eV have a 100 \% collection efficiency. The
averaged experimental electron flux was about 10$^{10}$-10$^{11}$
electrons cm$^{-2}$ s$^{-1}$.

A schematic diagram of both experimental setups is presented in
Fig.~\ref{fig:TOFcartoon}. In the upper and bottom figures the
vacuum chamber and the TOF-detector configuration for the electron
and proton impact experiments are shown respectively.

\section{Results and discussion}

Three formic acid electron impact mass spectra, obtained at 500,
1000 and 2000 eV impact energy are presented in Fig. 3a. The ionic
fragments are labeled and indicated by arrows in the first spectrum.
No significant changes are observed between the 0.5 - 2 keV electron
impact experiments. For the sake of comparison, in Fig. 3b three
mass spectra produced by proton impact with energies of 0.128, 1 and
2 MeV, respectively, are displayed. Again, only minor changes are
observed between these spectra. The main difference between the
proton and electron impact spectra is the peak broadening present in
the proton impact spectra which reflects mainly the different mass
resolution of the spectrometers.

\begin{figure}
 \centering
 \resizebox{\hsize}{!}{\includegraphics{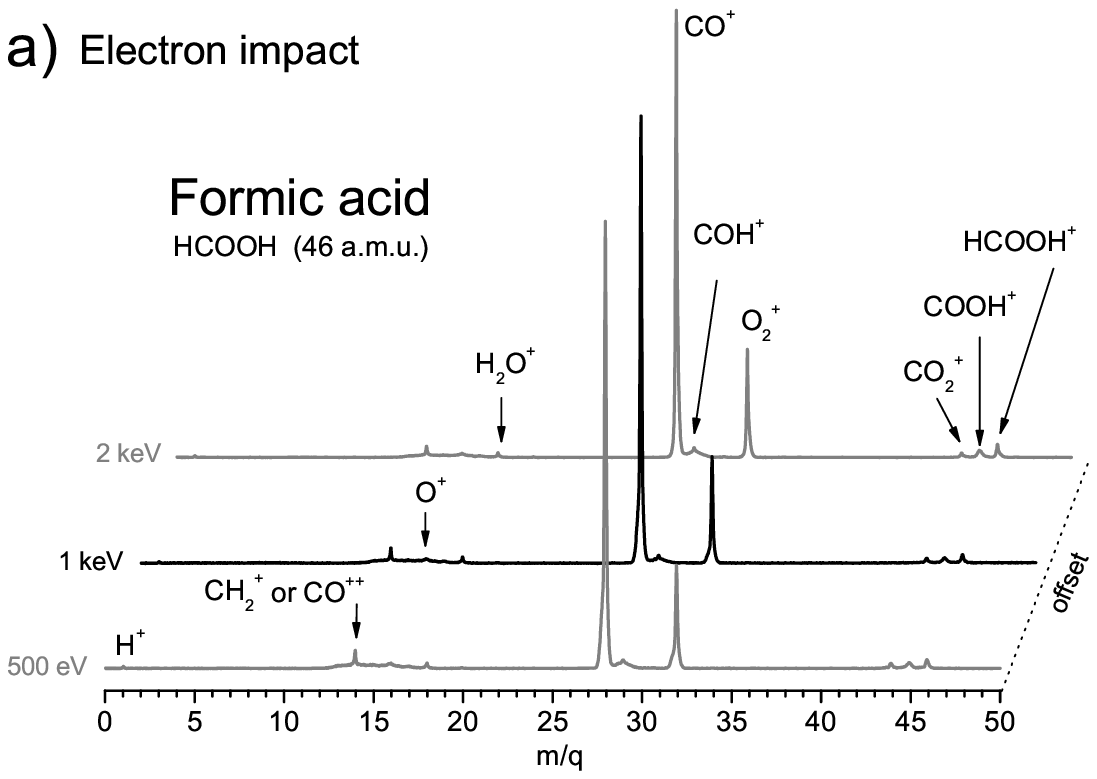}}
  \resizebox{\hsize}{!}{\includegraphics{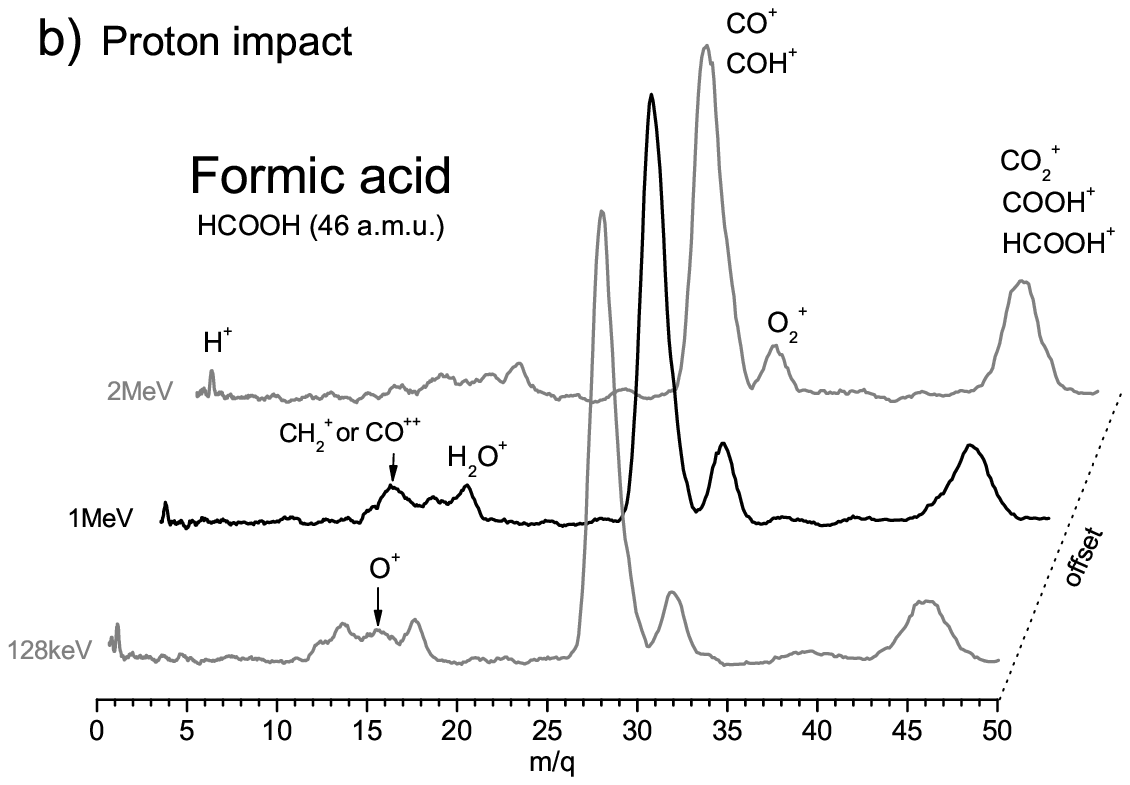}}
 \caption{Time-of-flight mass spectrum of HCOOH molecule. a) recorded at 500 eV, 1keV and 2 keV electron impact beam. b) recorded at 128 KeV, 1 MeV and 2 MeV proton impact beam.}
 \label{fig:ms-samples}
\end{figure}

\begin{figure}
\centering
 \resizebox{\hsize}{!}{\includegraphics{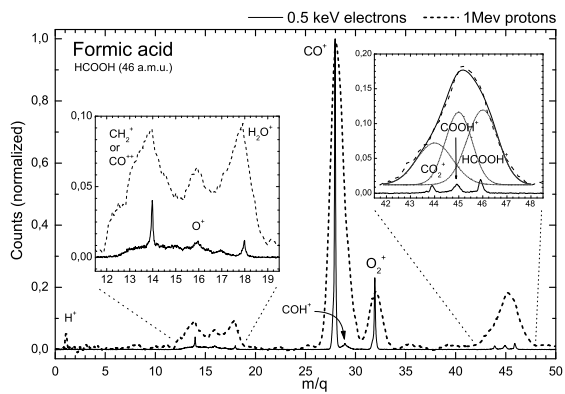}}
 \caption{Comparison between mass spectra of formic acid fragments by 1 MeV protons and
0.5 keV electrons.} \label{fig-comp}
\end{figure}

A direct comparison between time-of-flight spectra, expressed in
terms of mass over charge ratio, for the fragmentation of formic
acid by 1.0 MeV protons and by roughly equivelocity 0.5 keV electron
impact is presented in the Fig.~\ref{fig-comp}. Despite the fact
that both projectiles are leading to the same fragmentation pattern,
we can clearly observe differences in the relative yield of the
fragments (see Tables~\ref{tab:piy-eletr} and ~\ref{tab:piy-prot}).
The 0.5 keV electrons induce a stronger fragmentation of the formic
acid molecules than the equivelocity 1.0 MeV protons. This
observation is in direct accordance with the corresponding
differences noted also in the rare gas target-electron and proton
impact experiments (Melo et al. 2002) and has been suggested to
arise from projectile charge effects. In the inset Fig. 4 (at
right), we can see a deconvolution process of the highest mass peak
obtained in the proton case, showing the contributions of fragments
CO$_2^+$, COOH$^+$ and HCOOH$^+$ superimposed with the peaks
obtained in the electron case.

Since the electron and proton impact spectra have been obtained
under different mass resolution conditions we have grouped some
fragments (ex. CO group: CO$^+$ + COH$^+$) to avoid uncertainties in
the yield comparison. The fragments from each group are indicated by
the upper arrows. The most significant differences are found in the
yields of COO group (CO$_2^+$ + COOH$^+$ + HCOOH$^+$) and O$_2^+$.
The former present a yield 4 times higher than in the 1 MeV protons
experiment and the latter about 2 times higher than in the
experiment with 0.5 keV electrons.

Generally speaking, both projectile interactions lead to the same
formic acid fragments, but the overall relative intensities of the
fragments are clearly distinct for each incident projectile. In
addition, a comparison between both mass spectra conveys to some
interesting observations. For both projectiles, the CO$^+$ fragment
dominates the spectra, though, in the proton impact case, this peak
is clearly convoluted with the HCO$^+$ peak.

Another interesting observation concerns the relative intensity of
the H$^+$. In the proton case, it is roughly three times more
intense than in the electron impact case. Whether this is a result
of the spectrometer discrimination against high energetic fragments
(due to its low mass, H$^+$ takes most of the kinetic energy
available in the dissociation process), or associated with different
breakup mechanisms, is uncertain. On the other hand, the relative
intensity of the O$_2^+$ peak in the electron impact case overcomes
by a factor of 2 the corresponding peak in the proton impact case.

\subsection{Partial ion yield and kinetic energy release}

The Partial Ion Yield (PIY) or relative intensities for each
fragment $i$ was obtain by
\begin{equation}
PIY_i = \left( \frac{A_i }{A^{+}_t} \pm \frac{ \sqrt{A_i}+ A_i
\times ER/100}{A^{+}_t} \right) \times 100\%
\end{equation}
were $A_i$ is the area of a fragment peak, $A^{+}_t$ is the total
area of the PEPICO spectrum. The $ER$ = 2 \% (in the electron
impact) and 10 \% (in the proton impact) is the estimated error
factor due to the data treatment.

Considering that the electric field in the interaction region is
uniform, we can determine the mean kinetic energy released in the
fragmentation process ($U_0$) for each ionic fragment from the
correspodent peak width, as suggested by Simon et al. (1991), Hansen
et al. (1998) and Santos, Lucas \& de Souza (2001)

\begin{equation} \label{eq-U0}
U_0 = \Big(\frac{qE \Delta t}{2} \Big)^2 \frac{1}{2m}
\end{equation}

where $q$ is the ion fragment charge, $E$ the electric field in the
interaction region, $m$ is the mass of the fragment, and $\Delta t$
is the time peak width (FWHM) taken from time-of-flight mass
spectra.

The relative intensities and mean kinetic energy release $U_0$ (only
for the electron impact case) are presented in
tables~\ref{tab:piy-eletr} and \ref{tab:piy-prot}, as a function of
the incident energy for the electron and proton impact experiments,
respectively. The water peak (18 a.m.u.) represents an intrinsic
recombination of some dissociation products of HCOOH already
reported in the literature (Su et al 2000; Schwel et al 2002;
Boechat-Roberty, Pilling \& Santos 2005 and references therein).

\begin{table*}
\centering \caption{Relative intensities (PIY) and kinetic energy
$U_0$ release by the fragments in the formic acid mass spectra, as a
function of electron beam energy. Only fragments with intensity $>$
0.1 \% were tabulated. The estimated experimental error was below
10\%.} \label{tab:piy-eletr}
\begin{tabular}{ l l l r r r }
\hline \hline
\multicolumn{2}{c}{Fragments}           &  & \multicolumn{3}{c}{PIY (\%) / $U_0$ (eV)}\\
\cline{1-2}  \cline{4-6}
   $m/q$        & Attribution           &  & 500 eV        & 1000 eV         & 2000 eV       \\
\hline
1           & H$^+$                     &  & 0.79 / 1.5      & 0.47 / 1.1     & 0.76 / 0.81    \\
12, 13, 14  & C group$^a$               &  & 11.6 / ~~ -~~   & 10.0 / ~~ -~~  & 8.89 / ~~ -~~  \\
16      & O$^+$                         &  & 3.1 / 1.1       & 3.3 / 1.1      & 3.4 / 1.1    \\
17      & OH$^+$                        &  & 1.2 / 0.85      & 0.95 / 0.85    & 0.96 / 0.85    \\
18      & H$_2$O$^+$                    &  & 1.0 / 0.03      & 1.2 / 0.01     & 1.1 / 0.02    \\
28      & CO$^+$                        &  & 60.2 / 0.01     & 61.7 / 0.01    & 60.1 / 0.01    \\
29      & COH$^+$                       &  & 4.1 / 0.29      & 3.7 / 0.18     & 5.3 / 0.17    \\
32      & O$_2^+$                       &  & 14.5 / 0.01     & 15.0 / 0.01    & 14.7 / 0.01   \\
44      & CO$_2^+$                      &  & 0.83 / 0.02     & 0.92 / 0.01    & 0.74 / 0.01    \\
45      & COOH$^+$                      &  & 1.4 / 0.03      & 1.3 / 0.03     & 1.9 / 0.03    \\
46      & HCOOH$^+$                     &  & 1.3 / 0.01      & 1.4 / 0.01     & 2.0 / 0.01    \\
\hline \hline
\multicolumn{6}{l}{$^a$ C$^+$ + CH$^+$ + CH$_2^+$ or CO$^{++}$}\\
\end{tabular}
\end{table*}

\begin{table*}
\centering \caption{Relative intensities (PIY) of fragments in the
formic acid mass spectra, as a function of proton beam energy. Only
fragments with intensity $>$ 0.1 \% were tabulated. The estimated
experimental error was below 20\%.} \label{tab:piy-prot}
\begin{tabular}{ l l l r r r r r r }
\hline \hline
\multicolumn{2}{c}{Fragments}                 &  & \multicolumn{6}{c}{PIY (\%) }\\
\cline{1-2}  \cline{4-9}
   $m/q$        & Attribution                 &  & 128 keV & 200 keV & 500 keV & 1 MeV & 1.5 MeV & 2 MeV  \\
\hline
1             & H$^+$                         &  &  0.74   & 0.79   & 0.88   & 0.67   & 0.89     & 0.65  \\
12, 13, 14    & C group$^a$                   &  &  6.5    & 5.2    & 5.4    & 3.5    & 7.1      & 6.0   \\
16, 17, 18    & O group$^b$                   &  &  7.7    & 5.6    & 6.2    & 4.3    & 9.7      & 7.4   \\
28, 29        & CO group$^c$                  &  &  57.1   & 55.4   & 51.1   & 57.8   & 51.9     & 53.8    \\
32            & O$_2^+$                       &  &  9.1    & 7.4    & 8.2    & 12.3   & 7.9      & 6.7    \\
44, 45, 46    & COO group$^d$                 &  &  16.8   & 25.6   & 27.9   & 21.4   & 22.3     & 25.3    \\
44            & CO$_2^+ $                     &  &  3.7    & 10.0   & 9.8    & 5.6    & 6.4      & 5.8    \\
45            & COOH$^+ $                     &  &  7.9    & 9.4    & 11.2   & 8.7    & 8.0      & 13.0    \\
46            & HCOOH$^+ $                    &  &  5.2    & 6.5    & 6.9    & 7.1    & 7.9      & 6.5     \\
\hline \hline
\multicolumn{9}{l}{$^a$ C$^+$ + CH$^+$ + CH$_2^+$ or CO$^{++}$; $^b$ O + OH + H$_2$O; $^c$ CO$^+$ + COH$^+$; $^d$ COO$^+$ + COOH$^+$ + HCOOH$^+$  }  \\
\end{tabular}
\end{table*}

\begin{figure}
\resizebox{\hsize}{!}{\includegraphics{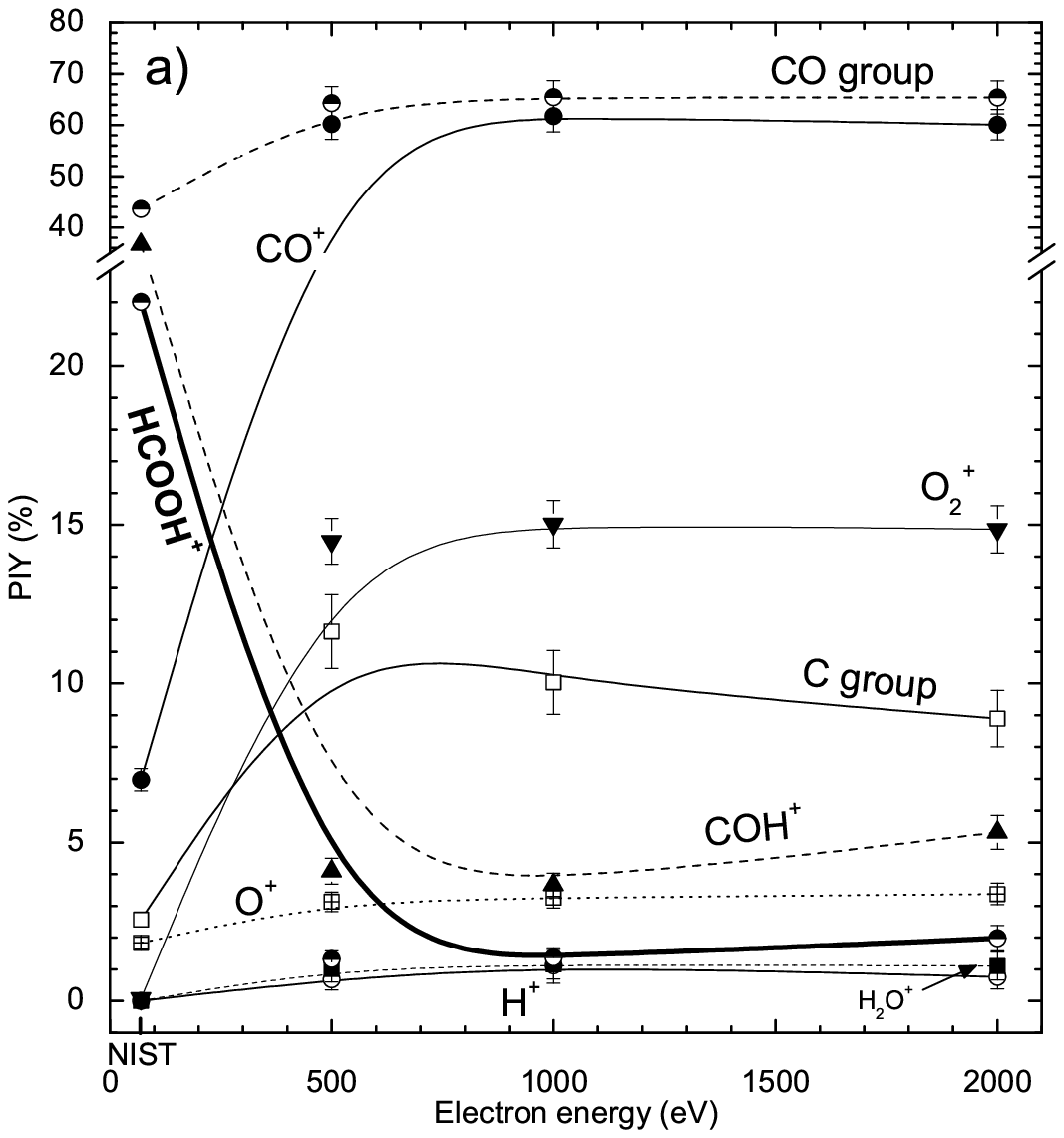}}
\resizebox{\hsize}{!}{\includegraphics{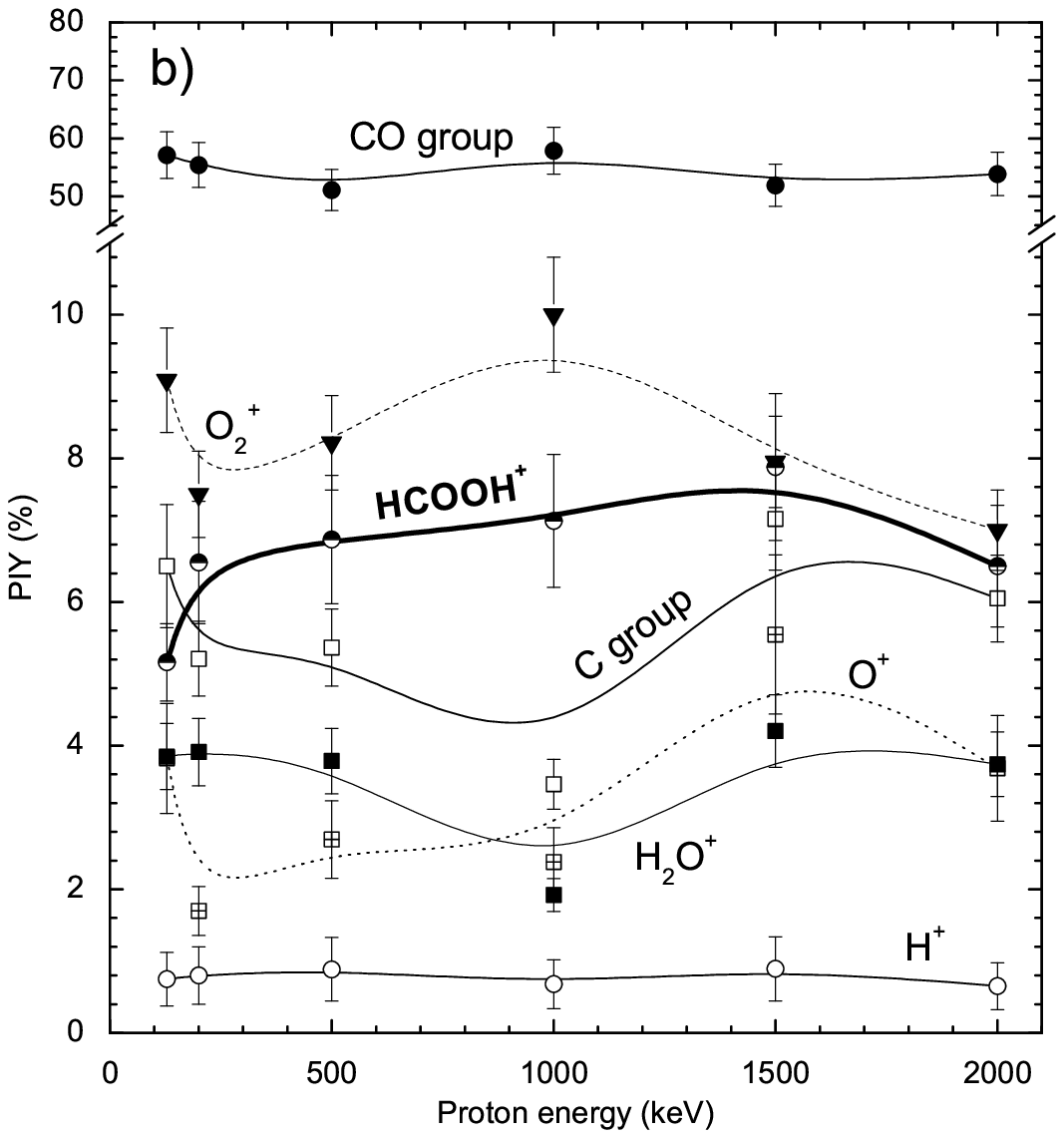}} \caption {a)
Comparison between partial ion yield (PIY) with 0.5 to 2 keV electon
beam and b) 128 keV to 2 MeV proton beam.} \label{fig:PIY-ELETePROT}
\end{figure}

The PIY for several fragments ejected in the dissociation process of
the formic acid molecule in the 500-2000 eV electron energy range
are displayed in Fig.~\ref{fig:PIY-ELETePROT}. The PIY for the 70 eV
electron impact data from NIST is also shown. The most abundant
fragments, CO$^+$ and O$_2^+$, exhibit almost a constant PIY in the
electron energy above 500 eV. The yields of the H$^+$, H$_2$O$^+$,
HCO$^+$, O$^+$ and HCOOH$^+$ molecules are included in the same
figure. The C group and the CO group are additionally shown, for
further comparison with the proton impact data. In the high energy
electron impact regime, we note that the PIY of the fragments
remains basically unaltered, as the electron beam energy increases,
with the exception of COH$^+$ and the C group, which present a small
variation with increasing projectile energy. The PIY obtained at
0.128-2 MeV proton impact for same fragments displayed in
Fig.~\ref{fig:PIY-ELETePROT}a, are shown in
Fig.~\ref{fig:PIY-ELETePROT}b. Despite some small fluctuations, the
PIY of the fragments does not show any specific trend as a function
of the proton energy.

In Fig.~\ref{fig:PIY-ELETePROT}b the PIY for 0.128-2 MeV proton
impact are shown for same fragments displayed in the
Fig.~\ref{fig:PIY-ELETePROT}a. Despite some small fluctuations the
PIY of the fragments does not indicate any specific tendency as
function of the proton energy.

\begin{figure}
\resizebox{\hsize}{!}{\includegraphics{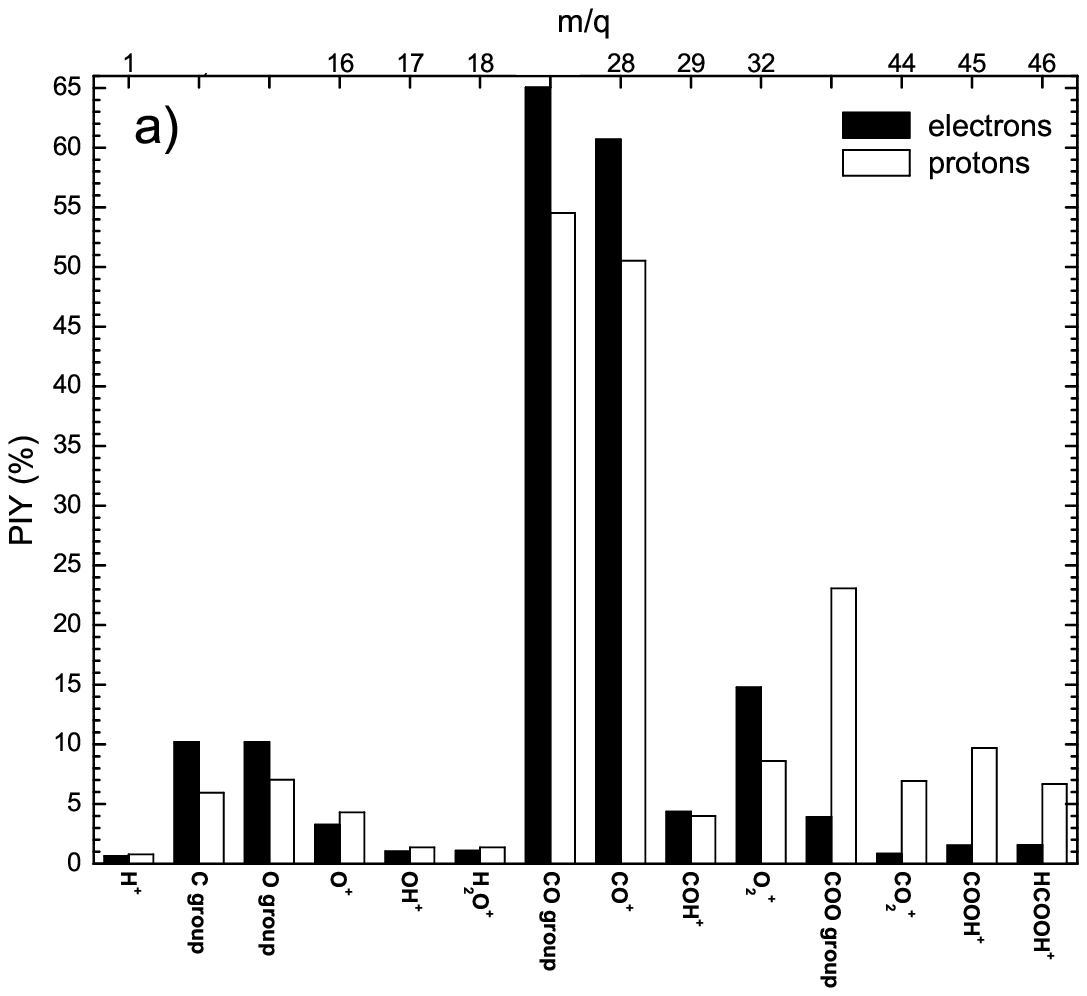}}
\resizebox{\hsize}{!}{\includegraphics{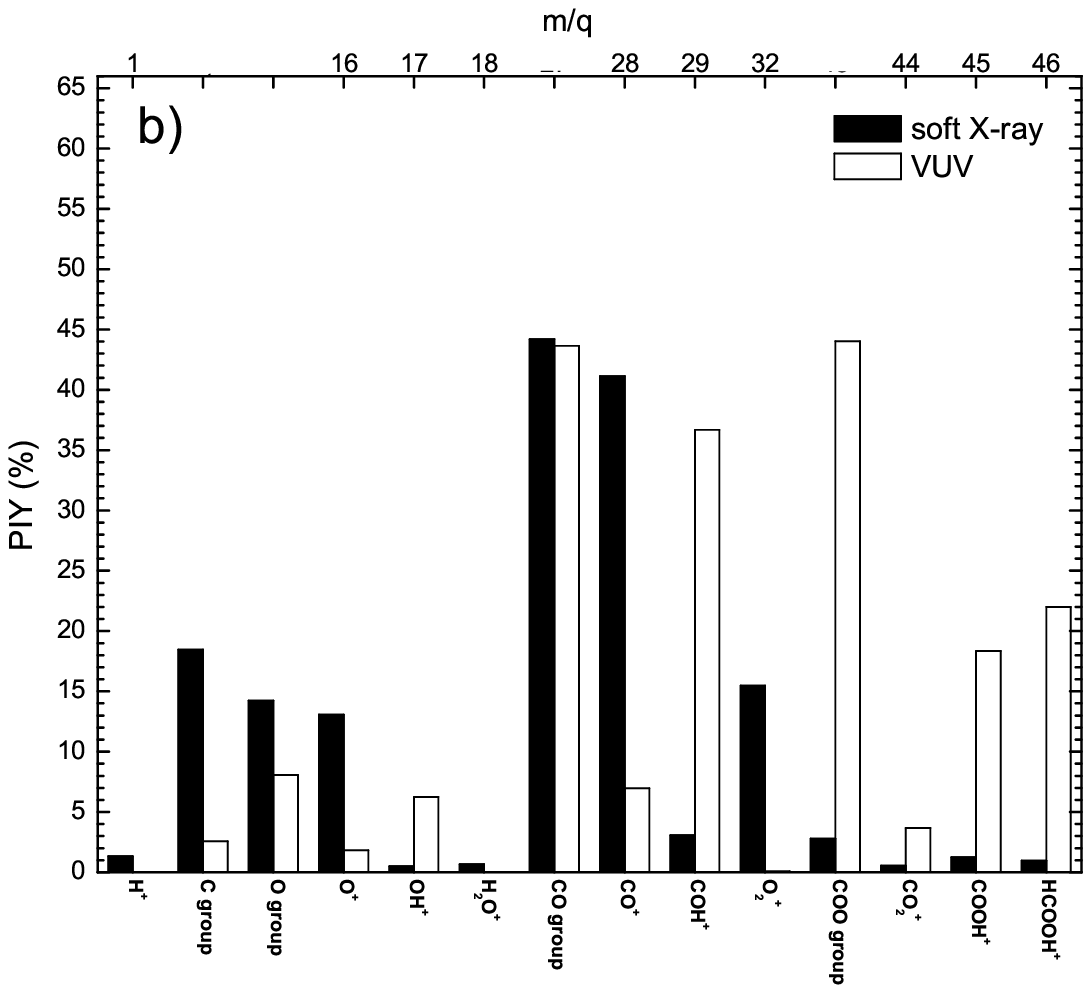}} \caption
{a) Comparison between partial ion yield (PIY) due dissociation with
electons bean (averaged between 0.5 to 2 keV) and a proton beam
(averaged between 128 keV to 2 MeV). b) Comparison between partial
ion yield (PIY) in soft X-ray field and VUV field from NIST 70 eV
electron impact mass spectra (adapted from Boechat-Roberty, Pilling
\& Santos 2005).} \label{fig:PIY-UVXRAY}
\end{figure}

A comparison between the averaged partial ion yield of the fragments
of HCOOH released by 0.5 to 2 keV electron and 0.128 to 2 MeV proton
beam impact could be seen in Fig.~\ref{fig:PIY-UVXRAY}a. The main
difference between the two impact regimens can be observed in the
yield of COO group fragments. In the case of proton impact the yield
of this group have shown roughly 5 times larger than the
corresponding yield associated with the electron impact process.
This corroborates the higher degree of dissociation promoted by the
energetic electron collisions.

Searching for possible similarities we present, in the
Fig.~\ref{fig:PIY-UVXRAY}b, a comparison between partial ion yield
of the fragments of HCOOH due to soft X-rays (290 eV) and VUV
``pseudo-photons'' (70 eV electron impact from NIST\footnote{The
dissociation induced by 70 eV electrons is very similar to the
dissociation induced by 21.21 eV (He I Lamp) photons. In both cases
the ionisation occurs in the valence shell (see discussion in Lago
et al 2004)}). The degree of destruction of the HCOOH molecule is
much larger in the soft X-rays case (Boechat-Roberty, Pilling \&
Santos 2005). Several fragments present a different pattern, as far
as the PIY are concerned, in X-ray field when compared to the UV
field. As an example, we could mention the large enhancement in the
production of CO$^+$ in the X Ray regime. The same is true for
O$^+$, O$_2^+$ and CH$_2^+$ fragments. On the other hand, the
opposite behavior is observed with the HCO$^+$ and OH$^+$ fragments,
which seem to be more efficiently produced by UV photons. This fact
has been previously reported (Suto et al. 1988, Su et al. 2000 and
Schwel et al. 2002).

The dissociation of the formic acid molecule by fast electrons and
by soft X-rays shows several similarities. The survival of the HCOOH
molecule, subjected to an incoming proton beam, was observed to be
6-8\% while in the case of the fast electrons was about 2\%, i. e.,
3 to 4 times smaller. The reactive COOH$^+$ fragment, also important
in pre-biotic ion-molecule chemistry, shows a similar behavior along
with the molecular ion HCOOH$^+$.

\subsection{Projectile momentum exchange}

It is accepted that the ionisation cross section, at high
velocities, are the same for electrons and protons (Bethe 1930). In
fact, within the first Born approximation, the total ionization
cross section $\sigma^{+}$ depends on the charge and the velocity of
the incoming projectile according to the relationship
\begin{equation}
\sigma^{+} \propto \frac{Z^2}{v^2} \ln{v}
\end{equation}
being fundamentally the same for equivelocity electrons and protons,
if the velocity is much larger than the binding energy of the active
electron (Fano 1954).

In addition, the projectile dependence enters also through the
$P_{min}$, the minimum momentum transfer. The value of $P_{max}$ can
be put to infinity without significantly affecting integrated cross
sections because the integrand decreases at large $P$. The minimum
momentum transfer occurs for scattering angles $\theta$, close to
zero (Rudd et al., 1985)
\begin{equation}
P_{min}^2 = \frac{\varepsilon^2}{v^2}  \Big( 1 +
\frac{\varepsilon^2}{2 E_p} + \ldots \Big)
\end{equation}
where $\varepsilon = IP + E_{ej}$ is the transferred energy, $IP$ is
the formic acid ionization potencial (11.33 eV), $E_{ej}$ is the
ejected electron energy, $v$ is the projectile velocity and $E_p$ is
projectile energy.

As most ejected electrons are formed with $E_{ej}$ close to zero,
for equivelocity 2.0 MeV protons and 1.0 keV electrons, the
calculated difference in the minimum momentum transfer is only 3~\%
larger for electrons, being virtually the same ($P_{min} \sim 0.047$
a.u.). As the projectile velocity decreases, the momentum transfer
difference increases, reaching 38 \% at $v$ = 2.26 a.u. (70 eV
electron impact or 128 keV proton impact). The difference can be
even higher due to the so-called trajectory effects: The Coulomb
interaction between the projectile and the target nuclei will tend
to accelerate and attract a negative projectile towards high
electron densities, whereas a positive particle will be decelerated
and repelled. In the case of proton impact, in addition, electron
capture by the projectile becomes an important process at impact
velocities close to the relevant target electron velocity further
suppressing the direct ionisation channel further, as can be seen in
Fig.~\ref{fig-sigma}c.

\subsection{Absolute ionisation and dissociation cross section}

The absolute cross sections for both ionisation and dissociation
processes of organic molecules by solar wind are extremely relevant
and essential as input for physico-chemical models (Haider \&
Bhardwaj 2005; Boice 2004; Rodgers et al. 2004) and molecular
abundance models (Sorrell 2001). In those theoretical models,
biomolecules are formed inside the bulk of icy grain mantles
photoprocessed by starlight (ultraviolet and soft X-rays photons)
and wind particles (protons, electrons, He$^+$). As mentioned by
Sorrell (2001), the principal uncertainty of these models comes from
the uncertainty of the dissociation cross section values
$\sigma_{d}$. Therefore the precise determination of $\sigma_{d}$ of
biomolecules is of paramount importance to estimate the molecular
abundance in comets and in the interstellar medium. Moreover,
knowing the particle flux and the $\sigma_{d}$ value we can estimate
the dissociation rate and the half-life of a specific molecule
(Cottin et al. 2003; Bernstein et al. 2004).

It is well known that the cross sections for a given collision
process related to the interaction of a fast projectile with a
molecular target can be estimated by summing up all the separate
cross sections of the constituent atoms of the molecule. Being
generally difficult to endorse on theoretical bases, the additivity
rule (i. e. the Bragg rule) is unremarkably the sole option
available in the lack of experimental ionisation cross section
values for molecular targets. Adoption of this procedure is grounded
on the premise that, for high projectile velocities, the target
molecule operates as an assembly of separate atoms being its
molecular nature negligible.

Watson et al. (2003) found that the total electron cross sections
measured with a variety of molecular targets, when divided by the
number of atoms per molecule and plotted versus target average
atomic number, closely mirrored the straight-line $Z^2$ dependence
established by the cross sections for the atomic targets He and Ne.
In the electron impact case, the additivity rule is also an
undefeated approach (Mark \& Dunn 1985; Bobeldijk et al. 1994;
Miller 1990;  Sun et al. 1998; Deutsch et al. 1996; Raj 1991;
Joshipura \& Patel 1994; Sun et al. 1994; Jiang, Sun \& Wan, 1995a
and 1995b) .

\begin{figure}
 \includegraphics[angle=0,scale=0.98]{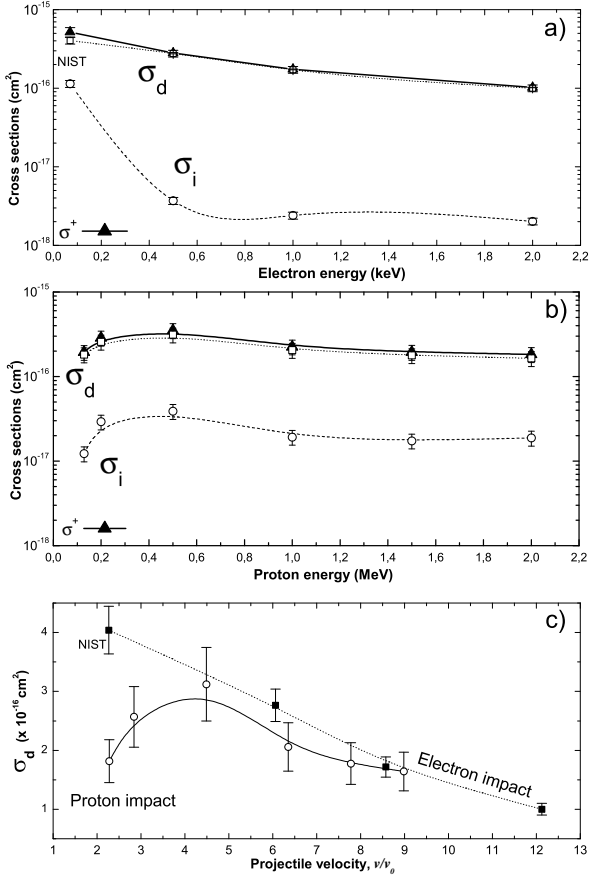}
\caption{a) Total ion production cross section ($\sigma^{+}$),
non-dissociative single ionisation cross section ($\sigma_{i}$) and
dissociative ionisation cross section ($\sigma_{d}$) of formic acid
as a function of electron beam energy. b) The same for proton beam
impact. c) Dissociation cross section of formic acid as a function
of electron impact velocity ($v/v_0 = \sqrt{E(eV)/13.6}$) and proton
impact velocity ($v/v_0 = 6.35 \sqrt{E(MeV)}$), for comparison. For
the cross section of the 70 eV electron impact on HCOOH we used the
PIY from NIST data base. See details in text.} \label{fig-sigma}
\end{figure}

The above mentioned observations mean that one can estimate the
formic acid total ionisation cross sections $ \sigma^+_{HCOOH}$ by
applying the additivity rule (ex. $ \sigma^+_{HCOOH}=
\sigma^+_{CO_2} + \sigma^+_{H_2}$). In the electron impact case, we
employed the averaged cross sections based on the theoretical
calculations (Kim \& Rudd 1994) for 70, 500, 1000 and 2000 eV
electrons on (O$_2$ + CH$_2$), (CO$_2$ + H$_2$), (CO + H$_2$O) and
(CH + HO$_2$). In the proton impact case, the cross section values
introduced for the normalization procedure are the averaged cross
sections based on the experimental determination (Rudd et al. 1985a
and 1985b) for 1.0 MeV and 2.0 MeV protons (CO$_2$ + H$_2$) and (CO
+ H$_2$O). The cross sections used for the normalisation in this
work are the average values of the summed pair of cross sections. In
both proton and electron cases the calculated uncertainties are
obtained from the chi-squared deviations.

The non-dissociative single ionisation (ionisation) cross section
$\sigma_{i}$ and the dissociative single ionisation (dissociation)
cross section $\sigma_{d}$ of formic by solar wind particles can be
determined by
\begin{equation}
\sigma_{i} = \sigma^{+} \frac{PIY_{HCOOH^+}}{100}
\end{equation}
and
\begin{equation}
\sigma_{d} = \sigma^{+} \Big( 1 - \frac{PIY_{HCOOH^+}}{100} \Big)
\end{equation}
where $\sigma^{+}$ is the cross section for single ionized fragments
of HCOOH  and $PIY_{HCOOH^+}$ is the partial yield of formic acid
ion in each impact case (see description in Boechat-Roberty, Pilling
\& Santos 2005).

Both cross sections are shown in Fig.~\ref{fig-sigma}, as a function
of the projectile energy. In the upper panel, Fig.~\ref{fig-sigma}a,
we display the total ionization cross section ($\sigma^{+}$), the
single ionization ($\sigma_{i}$) and the dissociation cross section
($\sigma_{d}$) of formic acid by energetic electron impact as a
function of electron energy. For the 70 eV electron impact cross
section we used the PIY from the NIST data base. In the middle
panel, Fig.~\ref{fig-sigma}b, we exhibit the equivalent set of cross
sections for the proton impact. A comparison of both dissociation
cross sections as a function of energetic particles impact velocity
($v/v_0 = \sqrt{E(eV)/13.6}$ for electrons; $v/v_0 = 6.35
\sqrt{E(MeV)/m(a.m.u.)}$ for ions) is presented in the lower panel,
Fig.~\ref{fig-sigma}c.

The experimental values for the total ion production cross section
($\sigma^{+}$), the non-dissociative single ionisation cross section
($\sigma_{i}$) and dissociative ionisation cross section
($\sigma_{d}$) of formic acid as a function of electron and proton
impact energies are presented in Table~\ref{tab-sigma}. The
estimated experimental error is considered to be lower than 30\%

\begin{table}
\centering \caption{Values of non-dissociative single ionization
cross section ($\sigma_{i}$) and dissociative ionization cross
section ($\sigma_{d}$) of formic acid as a function of electron
impact energy. The estimated experimental error was below 30\%.}
\label{tab-sigma}
\setlength{\tabcolsep}{4pt} %
\begin{tabular}{ l l c c c }
\\
\hline \hline
\\
Electron energy  (keV) &  & \multicolumn{3}{c}{Cross Sections (cm$^{2}$)} \\
\cline{3-5}   &  & $\sigma_{d}$ & $\sigma_{i}$ &
$\sigma^{+}$\\
\hline
0.07$^a$     &  & 4.04 $\times 10^{-16}$  & 1.14 $\times 10^{-16}$    & 5.18 $\times 10^{-16}$ \\
0.5          &  & 2.76 $\times 10^{-16}$  & 3.69 $\times 10^{-18}$    & 2.80 $\times 10^{-16}$ \\
1.0          &  & 1.72 $\times 10^{-16}$  & 2.40  $\times 10^{-18}$   & 1.74 $\times 10^{-16}$ \\
2.0          &  & 1.00 $\times 10^{-16}$  & 2.02 $\times 10^{-18}$    & 1.02 $\times 10^{-16}$ \\
\hline
\multicolumn{5}{l}{$^a$ PIY derived from NIST 70 eV electron impact mass spectra.}\\
\\
\\
Proton energy (MeV) &  & \multicolumn{3}{c}{Cross Sections (cm$^{2}$)} \\
\cline{3-5}   &  & $\sigma_{d}$ & $\sigma_{i}$ &
$\sigma^{+}$\\
\hline
0.128        &  & 1.82 $\times 10^{-16}$  & 1.23 $\times 10^{-17}$  & 1.94 $\times 10^{-16}$ \\
0.2          &  & 2.57 $\times 10^{-16}$  & 2.92 $\times 10^{-17}$  & 2.86 $\times 10^{-16}$ \\
0.5          &  & 3.13 $\times 10^{-16}$  & 3.89 $\times 10^{-17}$  & 3.51 $\times 10^{-16}$ \\
1.0          &  & 2.06 $\times 10^{-16}$  & 1.93 $\times 10^{-17}$ & 2.25 $\times 10^{-16}$ \\
1.5          &  & 1.78 $\times 10^{-16}$  & 1.74 $\times 10^{-17}$  & 1.95 $\times 10^{-16}$ \\
2.0          &  & 1.64 $\times 10^{-16}$  & 1.88 $\times 10^{-17}$  & 1.83 $\times 10^{-16}$ \\
\hline \hline
\\
\end{tabular}
\end{table}

\subsection{Dissociation rate and half-life}
The dissociation rate, $R$, of a molecule subjected to the
interstellar/interplanetary particle field (ex. protons or
electrons) in the energy range, $\varepsilon_2 - \varepsilon_1$, is
given by
\begin{equation} \label{eq-R}
R = \int_{\varepsilon_1 }^{\varepsilon_2} \sigma_{d}(\varepsilon)
I(\varepsilon) d\varepsilon
\end{equation}
where $\sigma_{d}(\varepsilon)$ is the dissociation cross section as
a function of projectile energy (cm$^2$) and $I(\varepsilon)$ is the
particle flux as a function of energy (particles cm$^{-2} eV^{-1}
s^{-1}$).

From Eq.~\ref{eq-R} we can also derive the half-life, $t_{1/2}$, of
the molecule as
\begin{equation}
t_{1/2} = \frac{\ln 2}{\int_{\varepsilon_1 }^{\varepsilon_2}
\sigma_{d}(\varepsilon) I_0(\varepsilon) d\varepsilon}
\end{equation}
which does not depend on the molecular number density.

A comparison between the dissociation rate and the half-life of the
formic acid due to the interaction with UV photons (taken from
literature) with the one determined from the interaction with
energetic solar wind electrons and protons (this work), are
presented in table~\ref{tab-rates}.

From Table 5, we can see that the half-life for HCOOH due to both
solar UV photons is about 1 order of magnitude lower than for the
energetic electrons. However, as we can see in Fig.
\ref{fig:particlesincomet}a, for the low energy electrons, the flux
is about several orders of magnitude higher than for the energetic
ones. This, combined with the fact that for small velocity electrons
the dissociation cross section present an increase (see Fig.
\ref{fig-sigma}a and c), may promote values even higher in the
dissociation rate for the small velocity electrons. These above
statements, points out also the importance of the electron impact
processes and data on chemical models.

Despite the similar fragmentation pattern in both cases, the
interaction with fast protons seems to be less significative on the
molecular cometary abundances. In fact, the determined half-life for
the formic acid due to proton impact is about 4 thousand years, 6
order of magnitude higher than for the electron impact case.

\begin{table}
\centering \caption{HCOOH Dissociation rate and half-life.}
\label{tab-rates}
\setlength{\tabcolsep}{2pt} %
\begin{tabular}{ l c c }
\hline \hline
Dissociation source                          & Dissociation rate (1/s)   & half-life (s)         \\
\hline
Quiet sun at 1 AU (UV photons) $^a$          & 8.8 $\times 10^{-4}$      & 7.9 $\times 10^{2}$   \\
Quiet sun at 1 AU (UV photons) $^b$          & 3.2 $\times 10^{-5}$      & 2.1 $\times 10^{4}$   \\
Lyman $\alpha$ $^c$                          & 3.9 $\times 10^{-6}$      & 1.7 $\times 10^{5}$   \\
Solar 0.07 - 2 keV electrons $^d$            & 9.8 $\times 10^{-7}$      & 7.1 $\times 10^{5}$   \\
Interstellar medium (theoretic) $^e$        & 6.7 $\times 10^{-10}$     & 1.0  $\times 10^{9}$  \\
Solar 0.1 - 2 MeV protons $^d$               & 5.7 $\times 10^{-12}$     & 1.2 $\times 10^{11}$  \\
 \hline
 \multicolumn{3}{l}{$^a$ Huebner et al. 1992; $^b$ Crovisier 1994; $^c$ Suto et al. 1988;}\\
 \multicolumn{3}{l}{$^d$ this work; $^e$ Roberge et al. 1991 (UV photons + cosmic rays).}
\end{tabular}
\end{table}

\section{Conclusions}

The effects of energetic charged particles (protons and electrons),
emanating from the solar wind, in the cometary gaseous formic acid
molecule (HCOOH) have been reproduced in our laboratory. Absolute
values for the ionisation and dissociation cross sections for the
interaction of this molecule with solar wind particles (electrons
and protons) have been determined. The energy range for the
energetic electrons was 0.5 keV to 2 keV, and from 128 keV to 2 MeV
for the fast protons.

In both cases mass spectra were obtained using time-of-flight mass
spectrometry and coincidence techniques. In agreement with previous
studies, the dissociation effects produced either by fast electrons
or by soft X-rays have presented several similarities
(Boechat-Roberty, Pilling \& Santos 2005). The survival of HCOOH
subjected to an incoming proton beam is found to be about 6-8\%
while in the fast electrons case it was about 2\%, i e, 3 to 4 times
smaller. The reactive COOH$^+$ fragment, also important in
pre-biotic ion-molecule chemistry, shows a behavior similar to the
observed with the molecular ion HCOOH$^+$ in both cases.

The average dissociation cross section due to both energetic
projectiles, in the energy range studied, is approximately 2~$\times
10^{-16}$ cm$^2$. However, a slight enhancement can be noted in the
case of fast electrons impact. The non-dissociative single
ionisation cross section present a value about 4-5 times higher in
the case of energetic proton which can be associated with the slight
enhancement in the total single ionization cross section due to
proton impact when compared with the energetic electrons. Moreover,
despite the fact that both projectiles lead to the same
fragmentation pattern, we observed differences in the relative
intensities of the ejected fragments.

The dissociation rates of formic acid due to UV photons and
energetic electrons roughly in the same range of magnitude, despite
the values for photons (from quite sun) it at least 30 times larger
than for fast electrons. This shows that, despite the interaction
with solar wind particles studied here seems to be quite secondary
compared to cometary photolytic processes, the electron impact
processes should be taken into account on chemical models in an
attempt to better simulate the interplanetary conditions. The above
statement become critical, for example, in case of solar
instabilities periods and if we consider the low energy ($<$ 10 eV)
electrons.

Given the fact that HCOOH is a minor cometary species, the studied
process may be have only a minor consequence on cometary chemistry,
for example, in the production of the extended sources of CO or
H$_2$CO. However, this could be not the case for a much more
abundant species such as methanol.

In the present work we have tried to contribute to the elucidation
of questions involving the extended molecular sources in the
cometary coma, such as the case of formaldehyde in the comet Halley
(Meier et al. 1993), frequently associated with the dissociation
process due to solar wind radiation. Based on our data, we expect
for example, that some of the detected cometary CO could be the
result of the dissociation processes of large molecules, as is the
case of the formic acid. We hope that the molecular cross section
derived in this work will give rise to more precise values for some
molecular abundances in cometary coma models.

%
\section*{Acknowledgments}
The authors would like to express their gratitude to MSc. F. C.
Pontes. This work was supported by the Brazilian funding agencies
FUJB (UFRJ), CAPES, CNPq, and FAPERJ.

%
%

\bsp

\label{lastpage}

\end{document}